\documentclass[12pt]{article}

\usepackage{amsmath}
\usepackage{amssymb}
\usepackage{scicite}
\usepackage{graphicx}
\usepackage{times}
\usepackage{rotating}
\usepackage{bm}

\newenvironment{sciabstract}{%
\begin{quote} \bf}
{\end{quote}}

\newcounter{lastnote}

\title{Spin Transfer Torques in MnSi at Ultra-low Current Densities}

\author
{F. Jonietz,$^{1}$ S. M\"uhlbauer,$^{1,2}$ C. Pfleiderer,$^{1\ast}$ A. Neubauer,$^{1}$\\
W. M\"unzer,$^{1}$ A. Bauer,$^{1}$ T. Adams,$^{1}$ R. Georgii,$^{1,2}$ P. B\"oni,$^{1}$ \\
R. A. Duine$^{3}$, K. Everschor$^{4}$, M. Garst$^{4}$, A. Rosch,$^{4}$\\
\\
\normalsize{$^{1}$Physik-Department E21, Technische Universit\"at
M\"unchen, D-85748 Garching, Germany}\\
\normalsize{$^{2}$Forschungsneutronenquelle Heinz Maier-Leibnitz (FRM II)}\\
\normalsize{Technische Universit\"at M\"unchen, D-85748 Garching, Germany}\\
\normalsize{$^{3}$Institute for Theoretical Physics, Utrecht University, 3584 CE Utrecht, The Netherlands}\\
\normalsize{$^{4}$Institute of Theoretical Physics, University of Cologne, D-50937 Cologne, Germany}\\
\normalsize{$^\ast$To whom correspondence should be addressed; E-mail:  christian.pfleiderer@frm2.tum.de.} }

\date{\today}

\begin{document}

\newcommand{\fcs}{Fe$_{1-x}$Co$_{x}$Si}

\baselineskip24pt \maketitle
\vspace{-12mm}
\begin{sciabstract}
Spin manipulation using electric currents is one of the most promising
directions in the field of spintronics.
We used neutron scattering to observe the influence of an
electric current on the magnetic structure in a bulk material.
In the skyrmion lattice of MnSi, where the spins form a lattice of magnetic vortices
similar to the vortex lattice in type II superconductors,
we observe the rotation of the diffraction pattern
in response to currents which are over five orders of magnitude smaller than those
typically applied in experimental studies on current-driven magnetization
dynamics in nanostructures. We attribute
our observations to an extremely efficient coupling of inhomogeneous spin
currents to topologically stable knots in spin structures.
\end{sciabstract}



\newpage The discovery of the effect of giant magnetoresistance,
now used commercially in hard disk drive industry,
is widely recognized as the starting point of the field of spintronics.
It represents the first example of electric
currents controlled efficiently by spin structures.
The complementary process of so-called spin transfer
torques, where magnetic structures and textures are manipulated by
electric currents \cite{slon96,berg96}, appears to be even more promising.
For instance, strong current pulses allow to
move ferromagnetic domain walls \cite{grol03,tsoi03}, switch magnetic
domains in multilayer devices \cite{tsoi98,myer99}, induce microwave
oscillations in nanomagnets \cite{kise03} and switch ferromagnetic
semiconductor structures \cite{yama04}. However, the typical current
densities required to create observable spin transfer torques
in present day studies exceed $10^{11}{\rm Am^{-2}}$.
Because this implies extreme ohmic heating it was generally believed that
spin torque effects can be studied exclusively in nanostructures.
We report the observation of spin transfer torques in a bulk material,
the skyrmion lattice phase of MnSi.
The spin transfer torques appear when the current density
exceeds an ultra-low threshold of $\sim10^{6}\,{\rm Am^{-2}}$ --
five orders of magnitude smaller than those
used typically in experimental studies on current-driven magnetization
dynamics in ferromagnetic metals and semiconductors.

\begin{figure}
\begin{center}
\includegraphics[width=0.65\textwidth]{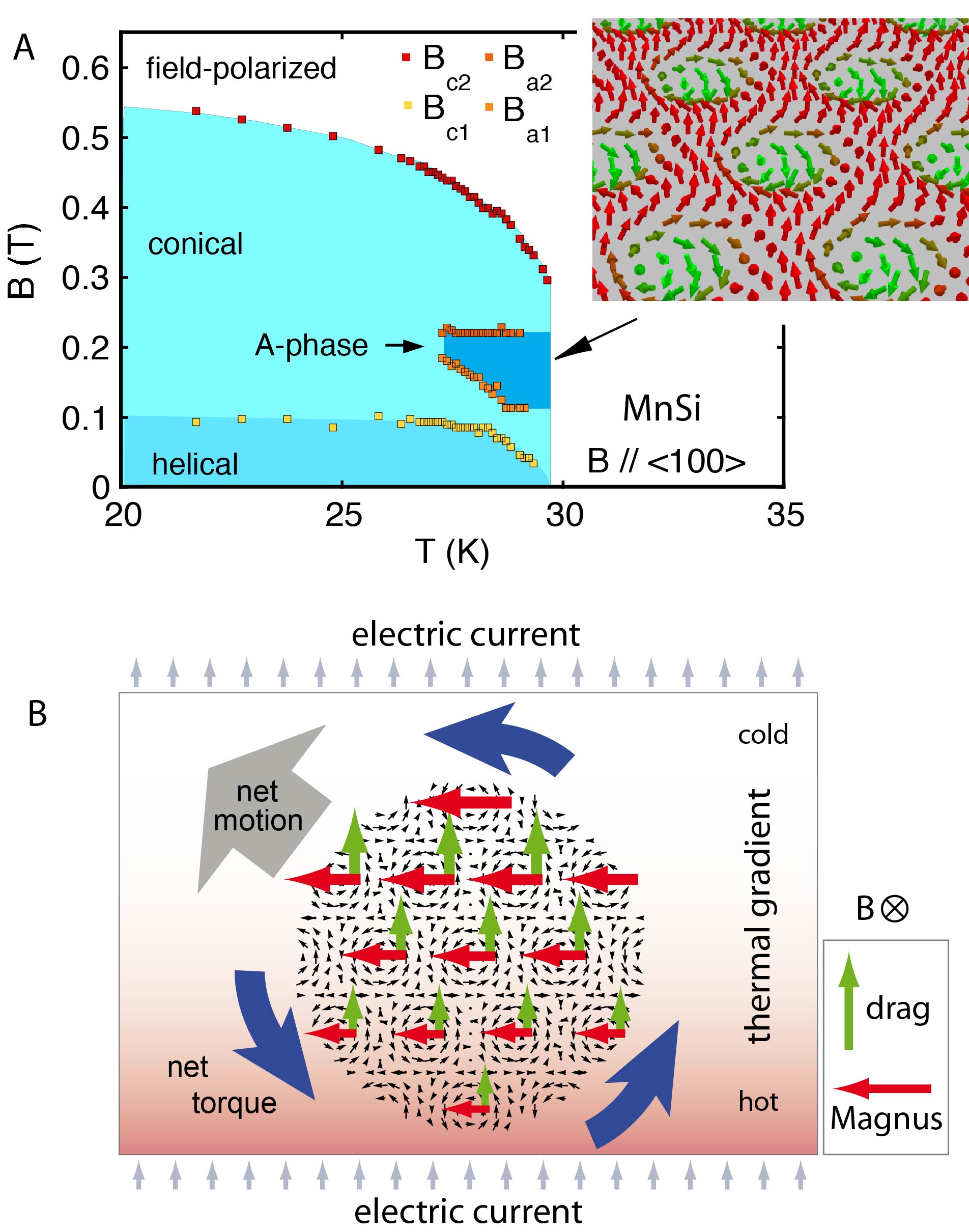}
\end{center}
\caption{(A) Magnetic phase diagram of MnSi. The inset schematically shows the
spin structure of the skyrmion lattice in a plane perpendicular to the applied field.
(B) Schematic depiction of the spin transfer torque effects on the skyrmion lattice. A temperature
gradient induces inhomogeneous Magnus and drag forces and therefore a rotational torque.
}
\label{figure1}
\end{figure}

The skyrmion lattice in chiral magnets, like MnSi and related B20 compounds,
was only recently discovered in neutron scattering studies
\cite{mueh09,muen09,neub09,pfleiderer:JPCM2010}
and confirmed to exist in Lorentz force microscopy for {\fcs} ($x=0.5$) \cite{yu:2010}.
It represents a new form of magnetic
order that may be viewed as a crystallization of topologically stable
knots of the spin structure that shares remarkable
similarities with the mixed state in type II superconductors.
For zero magnetic field (Fig.\,\ref{figure1}\,A) helimagnetic order appears
in MnSi below $T_c=29.5\,{\rm K}$.
In a small magnetic field the skyrmion lattice stabilizes in a
pocket below $T_c$, also known as the A phase. The spin structure of the
skyrmion lattice in MnSi consists of a hexagonal lattice of magnetic
vortex lines oriented parallel to the magnetic field $\bf B$
(inset to Fig.\,\ref{figure1}\,A).

Skyrmion lattices in chiral magnets are attractive for studies of spin
torque effects, because they are coupled very weakly to the atomic
crystal structure \cite{mueh09} and may be expected to pin very weakly
to disorder.  In addition, electric currents couple very
efficiently to skyrmions as follows.
When the conduction electrons in a metal move across a magnetic texture, their spin follows the local
magnetization adiabatically. Spins which change their orientation pick up a quantum mechanical phase, the Berry phase,
that may be viewed as an Aharonov-Bohm phase
arising from a fictitious effective field \cite{Ye:PRL1999,tata07,binz08}
${\bf B}^i_{\rm eff}=
\frac{\Phi_0}{8 \pi} \epsilon_{ijk} \hat{M} (\partial_j
\hat M \times \partial_k \hat M)$,
where $\hat M={\bf M}/|{\bf M}|$ is the direction of the local magnetization and
$\Phi_0=h/e$ is the flux quantum for a single electron.
In the skyrmion lattice $B_{\rm eff}$ has a topologically quantized
average strength of $-\Phi_0$ per area of the magnetic unit cell
(for MnSi $B_{\rm eff}\approx 2.5$\,T \cite{neub09}). $B_{\rm eff}$ induces an effective Lorentz force which gives rise
to an additional ``topological'' contribution to the Hall effect proportional to the product of $B_{\rm eff}$ and the local polarization
of the conduction electrons
as observed experimentally \cite{neub09,lee09}.
Correspondingly, because the electrons are deflected, a force is exerted
on the magnetic structure so that there is an efficient `gyromagnetic coupling' \cite{Thiele72} of the current
to the skyrmion lattice \cite{pfleiderer:Nature2010}.

From an alternative point of view,
the skyrmion lattice may be viewed as an array of circulating
dissipationless spin currents, because the skyrmions
are characterized by gradients in the spin-orientation
related to their quantized winding number.
This is analogous to superconductors, where dissipationless charge currents flow around quantized vortices
due to gradients of the phase. When an extra spin current is induced by driving an electric current
through the magnetic metal, the spin currents on one side of the skyrmion are enhanced while
they are reduced on the other side. As for a spinning tennis ball, this velocity difference
 gives rise to a Magnus force acting on the skyrmions. Note, however, that
 spin (due to spin-orbit coupling) is in contrast to charge
 not conserved and therefore this intuitive picture is incomplete. Most importantly, also further dissipative
 forces arise \cite{zhan04} which drag the skyrmions parallel to the current.

In Fig.\,\ref{figure1}\,B this magnetic Magnus force,
which is perpendicular to current and field
direction, is sketched together with the additional drag forces.
The Magnus and drag forces may lead to a translational
motion of the skyrmion lattice.
However, for the current densities used in our experiment the drift velocity of the electrons and therefore
also the drift velocity
of the skyrmions is not very large and thus
very difficult to detect in a scattering experiment.


In contrast to a translational motion, a rotation is much easier to measure in neutron scattering.
Thus, we performed our
experiment in the presence of a small temperature gradient parallel to the current,
causing the magnetization and therefore the spin currents to
vary in magnitude across a domain of the skyrmion lattice.
 In turn, the
strength of the Magnus force varies across the skyrmion lattice
(Fig.\,\ref{figure1}\,B), inducing a net torque. As estimated below, the torques are sufficiently
strong to induce rotations
which can be measured directly by neutron scattering.

\begin{sidewaysfigure}
\begin{center}
\includegraphics[width=0.5\textwidth,angle=-90]{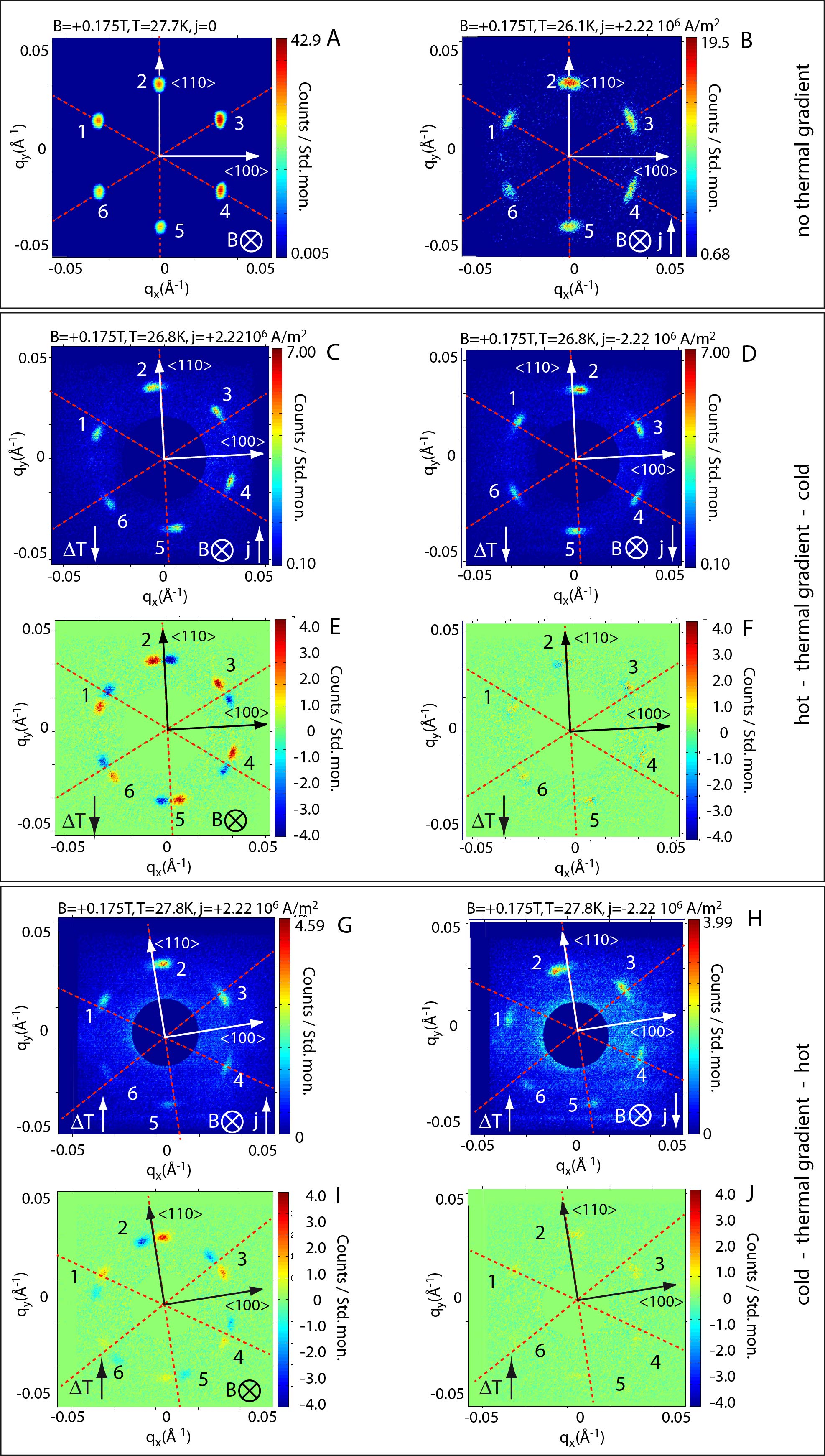}
\end{center}
\caption{Typical scattering intensity patterns observed in our
neutron scattering measurements  for a neutron beam parallel to the applied magnetic field \cite{SOM}. The red lines serve as a guide to the eye.
(A) Skyrmion lattice at zero current. (B) Pattern of A under current in the vertical direction (arrow).
 (C) When both the current and a small antiparallel  temperature gradient
  are present, the scattering pattern
 rotates counter-clockwise. (D) Pattern when reversing the current direction in C.
 (E) Difference between panels C and D. (F) Difference of intensities when reversing both current and field.
  (G) through (J): Same as panels C through F for reversed direction of the temperature gradient.
} \label{figure2}
\end{sidewaysfigure}

For our measurements an electric current was applied along bar-shaped
single crystals, where the direction of the current was always
perpendicular to the magnetic field and therefore to the skyrmion lines.
In the following the neutron beam was always collinear to the magnetic field \cite{SOM}.
The six-fold diffraction pattern of the skyrmion lattice at zero current,
$j=0$, (Fig.\,\ref{figure2}\,A) can be compared to the same scattering
pattern at a current density, $j=2.22 \cdot 10^6\,{\rm A\,m^{-2}}$,
first in a set-up minimizing any thermal gradients along the
current direction (Fig.\,\ref{figure2}\,B).
The current was applied along the vertical $[1\bar{1}0]$ direction,
whereas the field and the neutron beam were collinear to the line of sight and
along $[110]$ (the horizontal direction is along $[001]$).  Under current the
peaks of the diffraction spots remain in the same location and broaden
azimuthally.


We next generated a small temperature gradient along the direction of
the current as explained in Ref.\,\cite{SOM}.
Fig.\,\ref{figure2}\,C shows the diffraction pattern of the skyrmion
lattice for this set-up under an electrical current density,
$j=2.22\cdot 10^6\,{\rm A\,m^{-2}}$, which shows a pronounced
counter-clockwise rotation as compared to Figs.\,\ref{figure2}\,A and B
(note that arrows show the technical current direction and the line of sight is
opposite to the direction of the neutron beam consistent with convention).
When the current direction is reversed,
the rotation changes sign and the diffraction
pattern turns clockwise (Fig.\,\ref{figure2}\,D).

There are several unusual aspects of this rotation.
First, the entire scattering pattern rotates with respect to its center, i.e.,
all spots move by the same angle even though the electric current
has a distinct direction. Second, when reversing either the
direction of the current or the direction of the applied field
the sense of the rotation changes sign.
This is illustrated in Figs.\,\ref{figure2}\,E and F, which show the
difference of intensity under current reversal and
simultaneous reversal of current and field, respectively.
For the latter case the difference of intensities vanishes.

To confirm that the small temperature gradient along the current direction
causes the rotation of the scattering pattern,
we reversed the direction of the thermal gradient.
As illustrated in Fig.\,\ref{figure2}\,G and H this reverses the sense
of rotation with respect to the current and field direction
applied in Fig.\,\ref{figure2}\,C and D.
Thus the differences of intensity
under field reversal, shown in Fig.\,\ref{figure2}\,I, are reversed as compared
with Fig.\,\ref{figure2}\,E (red and blue spots have changed location).
When reversing both current and field direction, the difference of the patterns,
Fig.\,\ref{figure2}\,J, vanishes as before.
Finally, when applying
the current along a different crystallographic direction (we tested
$\langle111\rangle$ and an arbitrarily cut sample)
the same antisymmetric rotations of the diffraction pattern
as a function of magnetic field, electric
current and temperature gradient are observed \cite{SOM}.

\begin{figure}
\begin{center}
\includegraphics[width=0.6\textwidth]{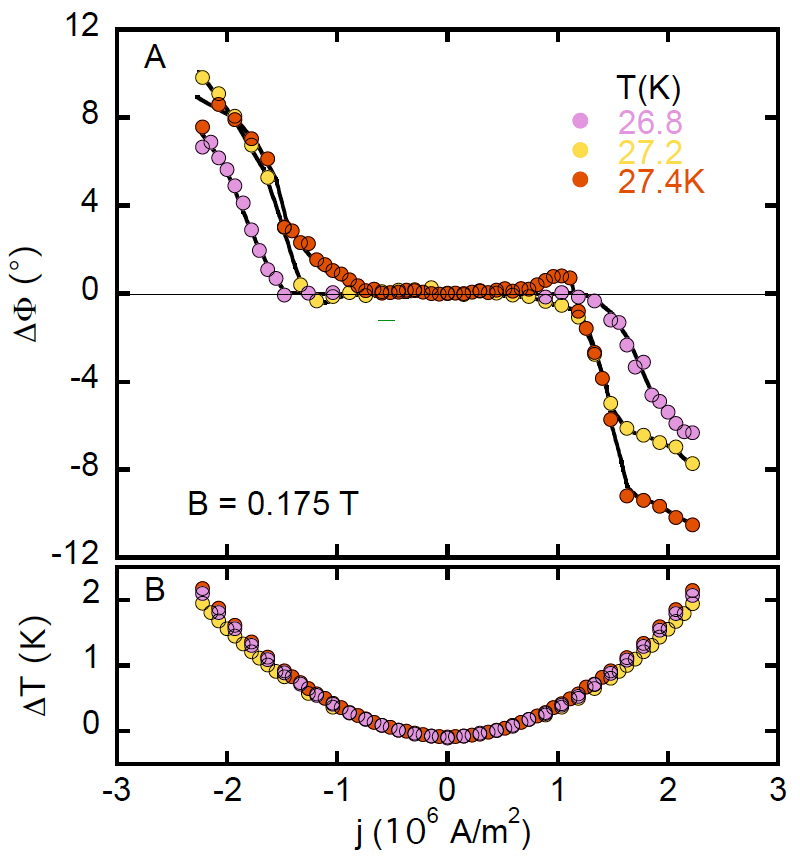}
\end{center}
\caption{(A) Change of the azimuthal angle of rotation of the scattering pattern as a function of
current density for three different temperatures. Data were recorded in a magnetic field of 0.175\,T. Above a clear threshold of $10^6\,{\rm A m^{-2}}$ an increasingly strong rotation
is observed, where the sign of the rotation depends on the direction of the current.
(B) Temperature difference between the surface of the sample and the sample holder as
a function of current density, where the temperature at the surface of the sample was kept constant for each of the three values given in Fig.\ref{figure3}A.
\label{figure3}}
\end{figure}

The detailed rotation as a function of the applied current
was determined in systematic measurements for various temperatures and samples.
Here the temperature measured at a specific spot at the surface
of the sample \cite{SOM} was kept constant and temperature gradients had always the same direction.
The resulting current dependence of the azimuthal rotation angle $\Delta\Phi$,
shown for three temperatures
as a function of the applied current density in Fig.\,\ref{figure3}, exhibits
a well-defined threshold of order $j_c\approx 10^{6}{\rm A\,m^{-2}}$
above which the rotation begins.
For $\vert j\vert>j_c$ the entire scattering pattern rotates and the
rotation angle $\Delta\Phi$ increases steeply with increasing $\vert j\vert-j_c$.

Potential parasitic effects cannot explain the observed rotations.
First, detailed studies show that there are no changes
of orientation of the skyrmion lattice as a
function of temperature at $j=0$
\cite{muen09,pfleiderer:JPCM2010,yu:2010}.
Second, the temperature difference
$\Delta T$ between the sample surface and the sample support
shows a smooth quadratic increase with current
density independent of the direction of the current (Fig.\,\ref{figure3}\,B)
in contrast to Fig.\,\ref{figure3}\,A.
Third, for the current
densities applied in our study the Oersted field increases from zero
(at the center of the sample) to a value of roughly 1\,mT at the
surface of the sample -- much smaller than the applied magnetic field
of 175\,mT and therefore negligible.
Finally, for a current parallel to the skyrmion lines or the pristine
helimagnetic state neither a rotation nor a broadening are observed.
Interestingly, recent numerical
simulations \cite{wess06,goto08} suggest that much larger current densities
of the order $10^{12}\,{\rm A\,m^{-2}}$ may change the orientation of
helical magnetic structures.

To explain our experiments the interplay of three tiny forces
has to be considered: (i) spin transfer torques, i.e., current induced forces,
(ii) pinning forces and (iii) anisotropy terms.
These determine the origin of the rotation,
the presence of a threshold and the angle $\Delta\Phi$ of the rotated
diffraction pattern, respectively.
The spin transfer torques can, e.g., be described by
a Landau-Lifshitz-Gilbert (LLG) equation or variants of Landau-Lifshitz Bloch
equations \cite{Schieback:PRB2009} which include both reactive and dissipative
components representing the Magnus and drag forces mentioned above, respectively.
Both are expected to be of the same order of magnitude \cite{he06}.
While the strength of the dissipative forces
(related to the parameter $\beta$ in the LLG equations) is not known,
it is possible to estimate the strength
of the reactive forces quantitatively.

The size of the Magnus force is given by the product of spin current and the fictitious effective magnetic field,
$f_M\approx e j_s B_{\rm eff}$,
and may be estimated as
 \begin{eqnarray}\label{size}
 {f}_M
 \approx  p(T)\cdot \frac{j}{10^6 {\rm A\,m^{-2}}} \frac{2.5 \cdot 10^6\, {\rm N}}{{\rm m}^3}
 \approx p(T)\cdot\frac{j}{10^6 {\rm A\,m^{-2}}}   \frac{2.7\cdot 10^{-10}   k_B T_c}{a^4}
\end{eqnarray}
where $a \approx 4.58$\,\AA\ is the lattice constant of MnSi,
$T_c\approx 29.5\,$K the ordering temperature, and the local temperature dependent
polarization is defined as the ratio of the spin and charge current densities times the
elementary charge, $p(T)=e j_s/j$.
For the skyrmion phase we estimate $p(T)\approx 0.1$~\cite{neub09}.
The resulting forces at the current densities of $10^6\, {\rm A\,m^{-2}}$ studied
are much larger than, e.g., gravitational forces on the sample,
but small when expressed in terms
of the microscopic units, $k_B T_c/a^4$ (cf. Eq.\,\ref{size}) raising the question
why the critical currents $j_c$ are so small.

For $j<j_c$ the current induced forces are balanced by pinning forces
caused by disorder and the underlying regular atomic crystal lattice.
The latter may be neglected because of the small spin-orbit interaction \cite{belitz06}.
The discussion presented in \cite{SOM} suggests,
that even a very strong defect,
which locally destroys the magnetization completely,
will result in a very small pinning force, less than a few
$10^{-5}\, k_B T_c/a$ per impurity, mainly because the magnetization of the skyrmions varies
very smoothly.
Therefore the observed critical current density $j_c$ together
with our estimate, Eq.\,\ref{size}, is consistent with strong pinning
defects with a density below 1\,ppm.
In fact, even though the real density of defects may be higher,
their influence may be strongly reduced as the system is in the `collective pinning' regime
known, e.g., from vortices in superconductors \cite{reviewVort94}.
Here pinning forces of random orientation average out to a large extent
due to the rigidity of the skyrmion lattice.

The size of the rotation of the skyrmion lattice for $j>j_c$
reflects the balance of the torques $\tau_M$ and $\tau_L$
due to inhomogeneous Magnus forces and the atomic lattice, respectively.
We start by noting, that by symmetry the orientation of a perfect skyrmion lattice
(described by a third rank tensor) cannot couple linearly to the current $j$,
because of the sixfold rotational symmetry of the lattice.
However, a small temperature gradient
breaks this symmetry and generates sizable variations of the
amplitude of the magnetization and associated polarization $p(T)$ of the electric currents,
because the skyrmion phase is only stable close to $T_c$ (cf. Fig.~\ref{figure1}).
As a result, the Magnus force, Eq.\,\ref{size}, which is proportional to the spin currents
and therefore to $p(T)$, will be considerably larger at the `cold' side of
the skyrmion lattice than at its `hot' side.
This gives rise to a net torque per volume
$\bm{\tau}_M \sim \int {\bf r} \times {\bf f}_M({\bf r}) d^3 r/V$.
Using Eq.\,\ref{size} with a spatially varying $p(T)$ and
ignoring again dissipative forces for an order of magnitude estimate,
we find that the rotational torque per volume, $\bm{\tau}_M$, in the direction of
the fictitious field $\bm{B}_{\rm eff}$ for a
skyrmion lattice domain of size $R$ is given by
\begin{eqnarray} \label{torque}
\tau_M  \sim 10^{-10} \frac{{\bf j} \cdot \bm{\nabla} p}{10^6 {\rm A\,m^{-2}}}
\frac{R^2}{a} \frac{k_B T_c}{a^3} \sim 10^{-5}
\frac{k_B T_c}{a^3} \left(\frac{R}{1\, {\rm mm}}\right)^2
\end{eqnarray}
where we assume $j\approx j_c$ and $\nabla p\approx 0.1/10\,{\rm mm}$

Note that the effect is proportional to the gradient of the
temperature parallel to the current as $\bm{\nabla} p\approx
\frac{\partial p}{\partial T} \bm{\nabla} T$ (Fig.\,\ref{figure1}B).
Thus $\bm{\tau}_M$ changes sign when either current, magnetic field or
temperature gradient are reversed. The sign of $\bm{\tau}_M$ and all
other forces obtained from this analysis (Fig.\,\ref{figure1}B) is
consistent with all our experiments taking into accout that charge
carriers are hole-like as measured experimentally \cite{neub09,lee09}. This
explains our main experimental results.  According to this analysis
the rotational torques arise from temperature-gradient induced
gradients in the spin-current.  Interestingly, it is more difficult to
generate an analogous effect for vortices in superconductors, because
charge, as opposed to the spin in skyrmion lattices, is exactly
conserved (in \cite{Lopez:PRL1999} a rotation of a superconducting
vortex lattice has been induced with a bespoke current distribution).

For an estimate of the factor $\left(R/1\, {\rm mm}\right)^2$ in Eq.~\ref{torque}
a lower limit, $R>1\mu{\rm m}$ may be inferred from the resolution-limited rocking
width of the magnetic Bragg peaks in the skyrmion phase
when avoiding demagnetization fields \cite{SOM}.
Yet, even a small torque $\tau_M$ may lead to large rotation
angles, because the balancing torque, $\tau_L$, which orients the
skyrmion lattice relative to the atomic lattice, is tiny. Only anisotropy terms  arising in
high power of the spin orbit coupling $\lambda_{SO}$ contribute to the torque per volume $\tau_L$ which we
estimate as \cite{SOM}
\begin{eqnarray}\label{counter}
\tau_L\sim - 10^{-2}\, \lambda_{SO}^4  \frac{k_B T_c}{a^3} \sin(6 \Phi).
\end{eqnarray}
For small rotation angles the torque $\tau_L$ grows linearly in the rotation angle $\Phi$.
However, in contrast to the torques arising from the inhomogeneous
Magnus force, $\tau_L$ is independent of the size $R$ of the domains.
The rotation angle $\Phi$ is finally determined by the balance of $\tau_M$ and $\tau_L$.
Because of the small prefactor in Eq.~\ref{counter}, $10^{-2} \lambda_{SO}^4$,
the large rotation angles observed in our experiments
can be explained even for moderately large domains.

It is likely that for $j>j_c$ not only a rotation by an angle sets in but also a linear motion of the magnetic structure, because any rotation of a sizable magnetic domain requires a depinning from defects. For moving domains, spin currents in a frame of reference that is comoving
with the domain enter in all formulas given above.
Therefore the size of the rotation in the end also depends
sensitively on the frictional forces which break Galilean invariance \cite{he06}.

Our observations identify chiral magnets
and systems with nontrivial topological properties as
ideal systems to advance the general understanding of the effects of
spin transfer torques. For instance, spin transfer torques may even be used to manipulate
individual skyrmions, recently observed directly in thin samples \cite{yu:2010}.
In fact, even complex magnetic structures at surfaces and interfaces
may be expected to exhibit the spin torque effects
we report here \cite{bode:Nature2007,acknow}.



\newpage

\newpage

\newpage

\newpage

\end{document}


\baselineskip24pt

\maketitle

\begin{sciabstract}
We present details of the experimental methods used in our studies of spin torque
 effects in the skyrmion lattice of MnSi. This concerns in particular the neutron scattering set up,
 sample environment, samples, data analysis and role of demagnetizing fields. We also extend the
 theoretical discussion of our results, discussing briefly pinning by impurities and the forces which determine the relative orientation
 of the skyrmion lattice and the MnSi lattice.
\end{sciabstract}

\newpage

\section{Experimental Methods}

\subsection{Neutron Scattering}
Our neutron scattering measurements were performed at the diffractometer MIRA at FRM II at the Technische Universit\"at M\"unchen \cite{geor07}. Data were recorded for an incident neutron wavelength $\lambda=9.6\,{\rm \AA}$ with a 5\% FWHM wavelength spread. A delayline $^3$He area detector of $200\times200\,{\rm mm^2}$ was used with a position resolution of order $2\times2\,{\rm mm^2}$.

The incident neutron beam was collimated over a distance of 1.5 m using apertures $4 \times 4$ mm$^2$ and 1.5 (width) $\times$ 3 (height) mm$^2$ after the monochromator and before the sample, respectively. The size of the sample was 1.5 (width) $\times$ 8 (height) mm$^2$, where we used a Cd aperture of $3\times3\,{\rm mm^2}$ directly at the sample in most experiments. The distance from the sample to the detector was between 0.8 and $1.3\,{\rm m}$.

\begin{figure}
\begin{center}
\includegraphics[width=0.7\textwidth]{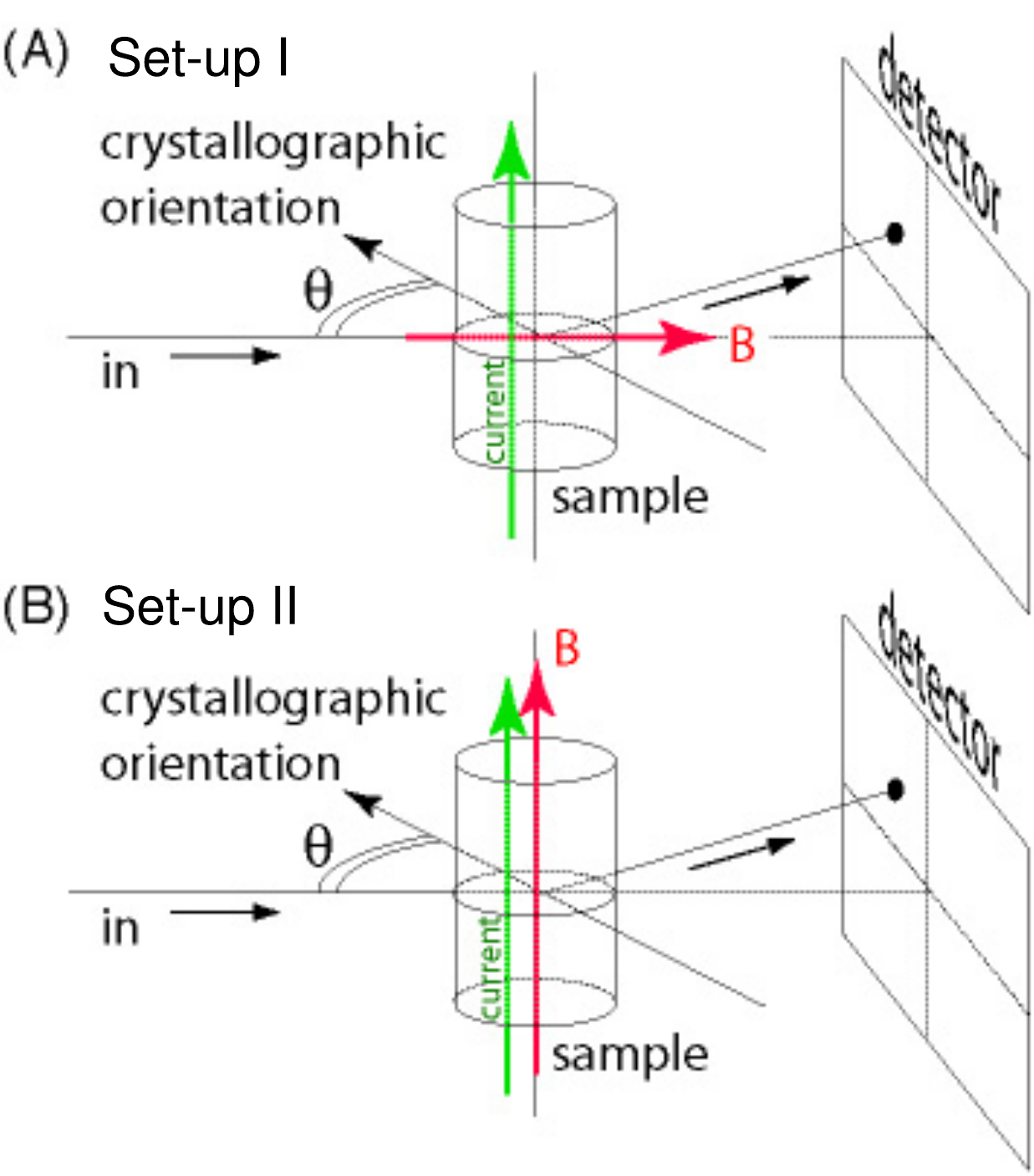}
\end{center}
\caption{(A) Set-up 1 used for all data shown in the main text. The incident neutron beam was parallel to the applied magnetic field (red arrow) that stabilized the skyrmion lattice, i.e. the neutron beam was parallel to the skyrmion lines. The applied electric current (green arrow) was perpendicular to the neutron beam and the magnetic field and hence skyrmion lines. (B) Set-up 2, in which the incident neutron beam was perpendicular to the applied magnetic field and the current. In this set up the current was flowing parallel to the skyrmion lines.} \label{figureS1}
\end{figure}

\subsection{Cryogenic environment and magnetic field}

Two different orientations between magnetic field, current direction and neutron beam were used. In set-up 1 the neutron beam was parallel to the applied magnetic field (red arrow) that stabilized the skyrmion lattice. The applied electric current (green arrow) was perpendicular to the neutron beam and the magnetic field. In set-up 1 the current was flowing perpendicular to the skyrmion lines. In set-up 2 the incident neutron beam was perpendicular to the applied magnetic field and the current. Therefore, in set-up 2 the current was flowing parallel to the skyrmion lines.

The magnetic field was generated with bespoke water-cooled Cu solenoids in a Helmholtz configuration \cite{mueh05}. The magnetic field profile was carefully characterized with a Hall probe and found to be uniform better than 1\% over the sample volume. In the A-phase data were identical after zero-field cooling and field cooling.

\subsection{Sample holder}

For our measurements we used a bespoke sample stick that permitted application of currents up to several 10\,A \cite{joni08}. The sample stick and sample holder were cooled with a cryogen-free pulse tube cooler, where He exchange gas with a pressure of several hundred mbar and spring-loaded mechanical contacts were used to ensure excellent thermal coupling.

\begin{figure}
\begin{center}
\includegraphics[width=0.9\textwidth]{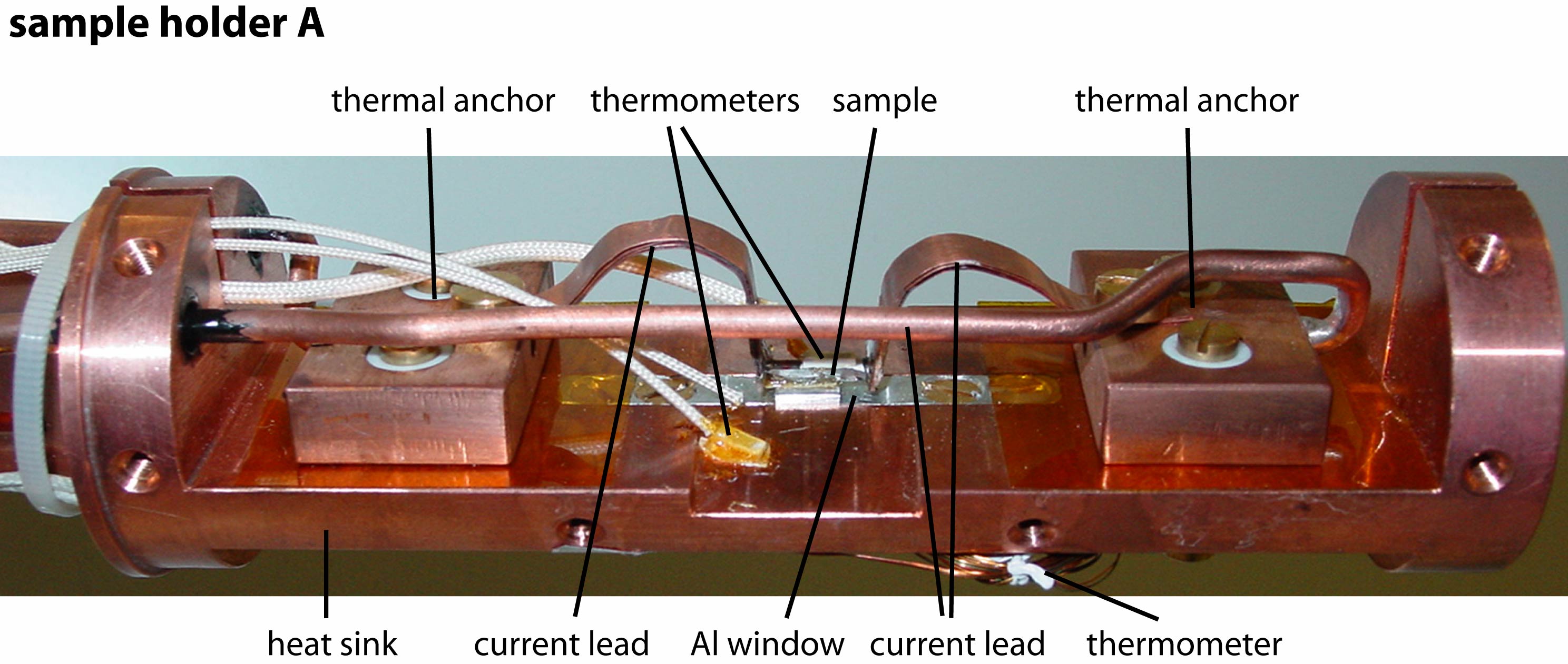}
\end{center}
\caption{Sample holder A used in our experiments. The sample and current leads were connected to a large Cu heat sink.  Data shown in the main text were recorded with this sample holder. The sample holder was cooled with a cryogen-free pulse tube cooler, where He exchange gas with a pressure of several hundred mbar and spring-loaded contacts were used for thermal coupling.} \label{figureS5}
\end{figure}

\begin{figure}
\begin{center}
\includegraphics[width=0.75\textwidth]{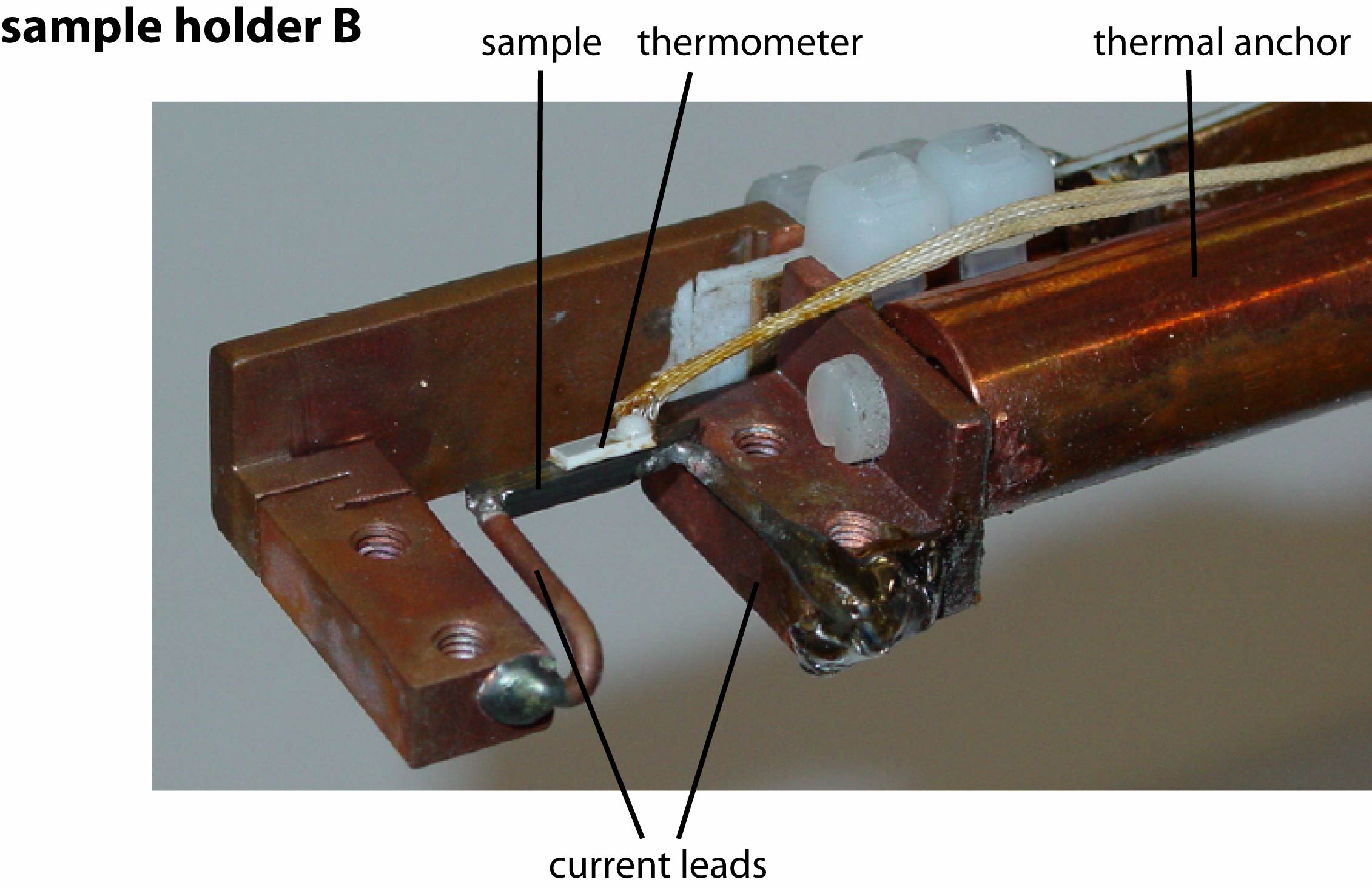}
\end{center}
\caption{Sample holder B used in our experiments. The sample holder was cooled with a croygen-free pulse tube cooler, where He exchange gas with a pressure of several hundred mbar and spring-loaded contacts were used for thermal coupling.} \label{figureS6}
\end{figure}

Two different sample holders were used as shown in Figs.\,\ref{figureS5} and \ref{figureS6}. The salient features reported in our paper were observed with both sample holders A and B. Sample holder A consisted of a large Cu heat sink, to which the sample and current leads were attached. Most of our data were measured with sample holder A, in particular all the data shown in the main text.

To insulate the sample electrically from the heat sink we used cigarette paper and a thin layer of GE varnish. In the low temperature section of the sample stick and sample holder the current leads were made of electrolytic (high-purity) Cu, soft-soldered directly to the samples with the help of special solder flux.

We generated a small temperature gradient along the direction of
current flow with a small wedge of GE varnish between the sample and
the sample holder.  This way the thermal coupling to the sample holder
changed along the sample, causing a temperature gradient along
the direction of the electric current in the presence of the resistive
heating by the sample.
The tiny tilt of the sample caused by the GE varnish
has no observable effect as we checked explicitly.

In both sample holders a Pt sensor was attached to the sample surface (the sensor did not generate any noticeable background signal in neutron scattering). Additional temperature sensors, e.g., as shown in Fig.\,\ref{figureS5}, were used to monitor the temperature difference between sample surface and heat sink (main body of the sample holder).

Typical temperature differences between the surface of the sample and the heat sink for the case of sample holder A are shown in Fig.\,3\,(B) of the main text. These data were obtained for fixed sample temperature while the temperature of the heat sink was adjusted. With increasing current density the temperature of the sample increases as the square of the current density, i.e., the increase is very smooth. This smooth increase contrasts the abrupt onset of the rotation above the critical current density $j_c$ as discussed in the main text.

\begin{figure}
\begin{center}
\includegraphics[width=0.7\textwidth]{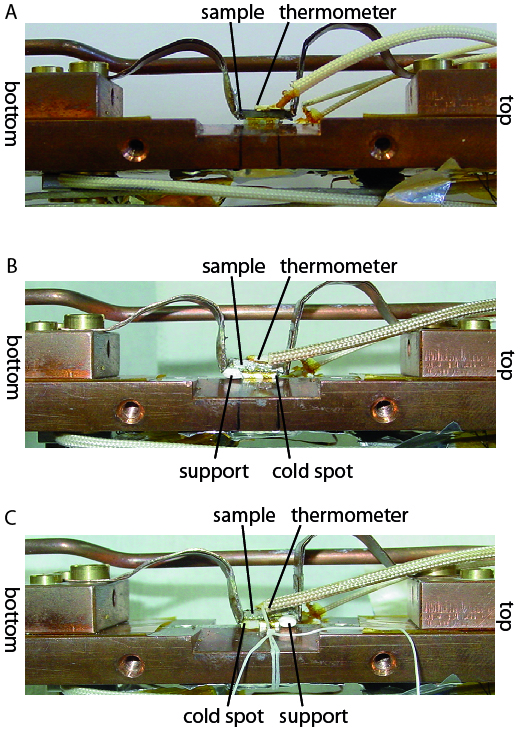}
\end{center}
\caption{(A) Sample (i) as firmly attached to the heat sink. Data shown in Fig.\,2\,(A) and 2\,(B) of the main text were recorded with this configuration. (B) Sample (i) as attached to the heat sink with a small wedge of GE varnish. To obtain a well-defined angle a small support (white) was placed underneath the sample. The cold spot is at the right hand side, which corresponds to the top of Fig.\,2 of the main text. Data shown in Fig.\,2\,(C) through (F) of the main text were recorded with this configuration. (C) Sample (i) as attached to the heat sink, where the wedge of GE varnish is inverted. The cold spot is at the left hand side, which corresponds to the bottom of Fig.\,2 of the main text. Data shown in Fig.\,2\,(G) through (J) of the main text were recorded with this configuration. } \label{figureS8}
\end{figure}

Fig.\,\ref{figureS8}\,(A) shows sample (i) as firmly attached to the heat sink. Data shown in Fig.\,2\,(B) of the main text were recorded with this configuration. Fig.\,\ref{figureS8}\,(B) shows sample (i) as attached to the heat sink with a small wedge of GE varnish. To obtain a well-defined angle a small support (white) was placed underneath the sample. The cold spot is at the right hand side, which corresponds to the top of Fig.\,2 of the main text (thermal gradient pointing down). Data shown in Fig.\,2\,(C) through (F) of the main text were recorded with this configuration. Fig.\,\ref{figureS8}\,(C) shows sample (i) as attached to the heat sink, where the direction of the wedge of GE varnish is reversed. The cold spot is at the left hand side, which corresponds to the bottom of Fig.\,2 of the main text (thermal gradient pointing up). Data shown in Fig.\,2\,(G) through (J) of the main text were recorded with this configuration.

\subsection{Samples}

We studied six different samples as summarized in Table\,\ref{table-samples}. The samples were either prepared by Bridgman growth or optical float-zoning. All samples studied were single crystals with high residual resistance ratios (RRR) around 100 and residual resistivity around $\rho_0\approx 2\,{\rm \mu\Omega cm}$. This corresponds to mean free paths up to 1000\,{\AA} \cite{brown-thesis}.
The magnetic and thermodynamic properties of samples cut from the same ingots agreed very well with the literature, i.e., all samples were of high quality.

\begin{figure}
\begin{center}
\includegraphics[width=0.8\textwidth]{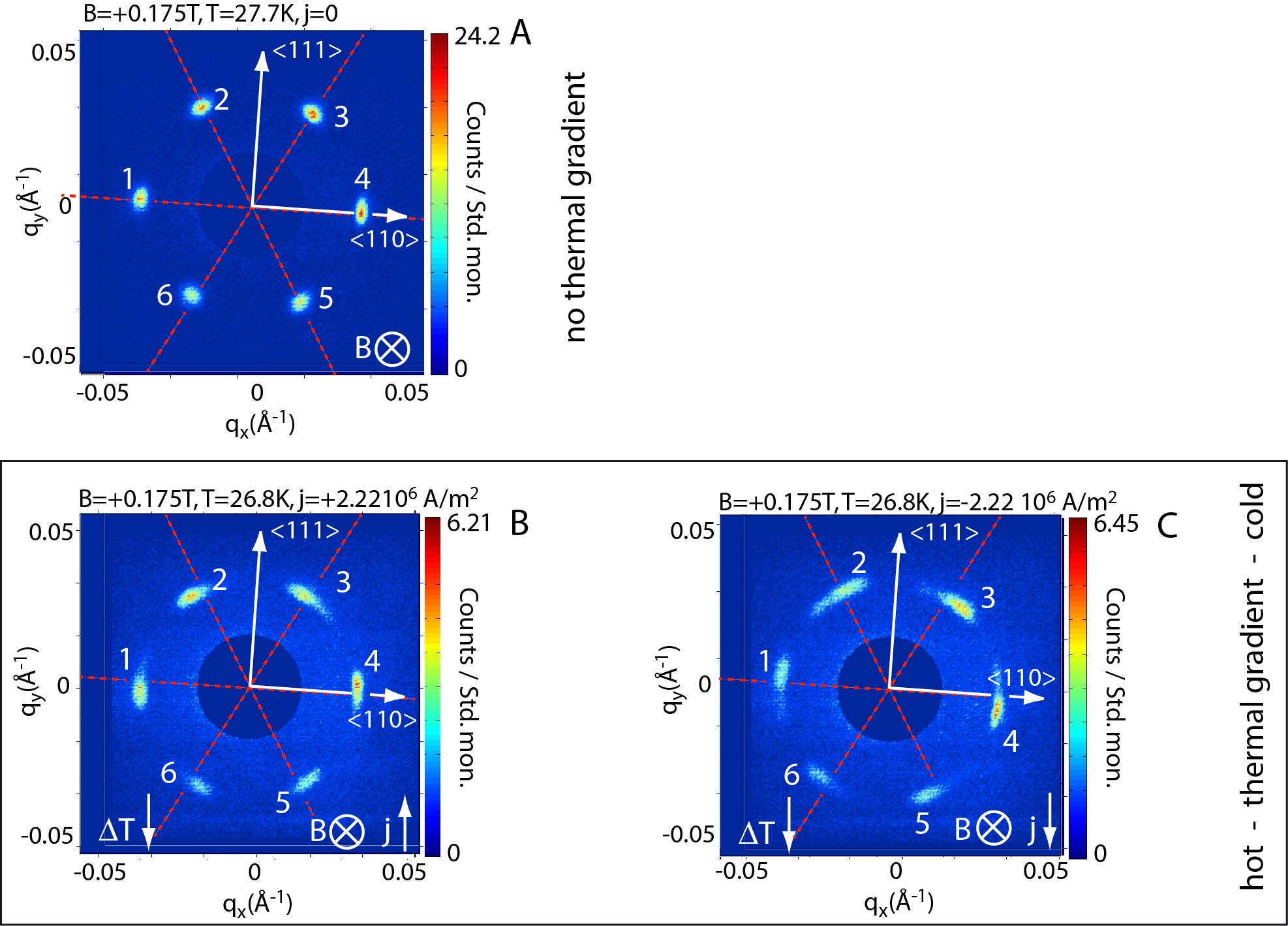}
\end{center}
\caption{(A) Scattering pattern of sample (vi) without current and hence temperature gradient. (B) Scattering pattern of sample (vi) under current and small temperature gradient. (C) same as panel B with current direction reversed.} \label{figureS10}
\end{figure}

For the experiments the samples were oriented with Laue x-ray diffraction. Note that data for the same samples has also been reported in the other publications, e.g., sample (ii) corresponds to "sample 2" in Ref.\,\cite{mueh09}. We note in particular, that sample (v) exclusively served to study the role of demagnetizing fields as described below. All samples showed the same salient properties under applied currents. This includes in particular sample (iii) which had a different aspect ratio of the cross-section and sample (vi), in which the current was applied along $\langle111\rangle$. Typical data recorded in sample (vi) are shown in Fig.\,\ref{figureS10}. In other words, the results reported in the main text do not depend on the specific crystallographic orientation and sample shape.

\newpage
\begin{table}
\centering
\caption{
MnSi single crystals investigated in this study.  'Length' refers to the size perpendicular (vertical) to the beam direction and represents the direction of current flow; 'thickness' refers to the size in the direction of the neutron beam; 'width' refers to the size perpendicular (horizontal) to the beam direction.
}
\vspace{5mm}
\label{table-samples}
\begin{tabular}{lllllll}
\hline\noalign{\smallskip}
 & length & thickness & width \\
\noalign{\smallskip}\hline\noalign{\smallskip}
sample (i) & $\sim$10\,mm & 1.5\,mm & 1.8\,mm	 \\
 & $[1\bar{1}0]$ & $[001]$ & $[110]$ \\
\noalign{\smallskip}\hline\noalign{\smallskip}
sample (ii) & $\sim$10\,mm & 	1.7\,mm & 1.9\,mm \\
 & $[1\bar{1}0]$ & $[001]$ & $[110]$ \\
\noalign{\smallskip}\hline\noalign{\smallskip}
sample (iii) & $\sim$10\,mm & 0.5\,mm & 4\,mm \\
 & $[1\bar{1}0]$ & $[001]$ & $[110]$ \\
\noalign{\smallskip}\hline\noalign{\smallskip}
sample (iv) & 12\,mm & 1.4\,mm & 1.95\,mm  \\
not oriented single crystal  \\
\noalign{\smallskip}\hline\noalign{\smallskip}
sample (v) & 8\,mm & 15\,mm & 0.8\,mm \\
 & $[1\bar{1}0]$ & $[001]$ & $[110]$ \\
\noalign{\smallskip}\hline\noalign{\smallskip}
sample (vi) & $\sim$5\,mm & 1.5\,mm & 1.8\,mm \\
 & $[1\bar{1}1]$ & $[1 \bar{1} \bar{2}]$ & $[110]$ \\
\noalign{\smallskip}\hline
\end{tabular}
\end{table} 

\subsection{Data for currents applied parallel to the Skyrmion lattice}

\begin{figure}
\begin{center}
\includegraphics[width=0.75\textwidth]{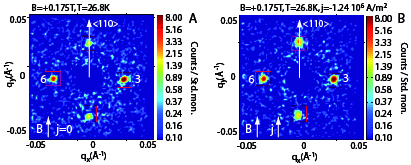}
\end{center}
\caption{(A) Diffraction pattern in the skyrmion lattice phase for magnetic field perpendicular to the incident neutron beam. Data were recorded in sample (ii). The spots labelled as '3' and '6' correspond to the same spots shown in Fig.\,2 of the main text for magnetic field parallel to the neutron beam. (B) Diffraction pattern in the skyrmion lattice for electric current parallel to the skyrmion lines. Note that the spots of the skyrmion lattice do not show any rotation or broadening whatsoever. For the same current density applied perpendicular to the skyrmion lines strong rotations are seen.} \label{figureS9}
\end{figure}

Typical data of the skyrmion lattice for the applied magnetic field perpendicular to the incident neutron beam (set-up II in Fig.\,\ref{figureS1}) are shown in Fig.\,\ref{figureS9}. Data were recorded for sample (ii), which showed a slightly lower critical current than sample (i). Data were recorded with sample holder B, where Fig.\,\ref{figureS9}\,(A) shows the scattering pattern for $j=0$.

Due to a small temperature gradient that exists in sample holder B at $j=0$ there are weak diffraction spots of the conical phase (spots at bottom and top) even at $j=0$. The horizontal spots labelled '3' and '6' correspond to spots '3' and '6' of the six-fold patterns shown in Fig.\,2 of the main text (data in Fig.\,2 were recorded in sample (i)).  The diffraction pattern under an applied electric current density of $j=1.24\cdot10^6\,{\rm A/m^2}$ \textit{parallel} to the skyrmion lines is shown in Fig.\,\ref{figureS9}\,(B). Neither a rotation nor broadening may be seen. Note that the same sample (number (ii)) as studied on the same sample holder shows strong rotations and broadening when the same current is applied perpendicular to the skyrmion lines just like the features reported in the main text (cf. Fig.\,2).

\subsection{Data Analysis}
Neutron data were analyzed with the software GRASP, v4.26, developed at the ILL. For each data set recorded at fixed current and temperature we determined at first accurately the center of the scattering pattern. We found slight systematic variations of the center of the scattering pattern as a function of current density. All azimuthal positions and widths of the intensity maxima reported in our manuscript were determined with respect to the center of the pattern. Data were binned as a function of azimuthal angle in intervals of 1 degree in a ring with respect to the center of the scattering pattern encompassing the scattering maxima. The azimuthal position and width was determined by fitting a simple Gaussian function.


\subsection{Role of Demagnetizing Fields}
\label{demag}

The mosaic spread of the skyrmion lattice was inferred from rocking scans. In some of the experiments we could only perform rocking scans with respect to a vertical axis. As a result the spots at the top and bottom may be much weaker because they do not satisfy the scattering condition.

In this section we summarize a recent study, where we compared the apparent mosaic spread derived from the rocking width of the magnetic satellites in the skyrmion lattice for different sample cross-sections. A full account of this work will be reported elsewhere \cite{adam09a}. In our study we find that the hexagonal magnetic scattering intensity in the A phase aligns strictly perpendicular to the applied magnetic field. This implies that the magnetic scattering intensity depends sensitively on variations of the magnetic field directions across the sample volume, notably the effects of demagnetizing fields.

\begin{figure}
\begin{center}
\includegraphics[width=0.6\textwidth]{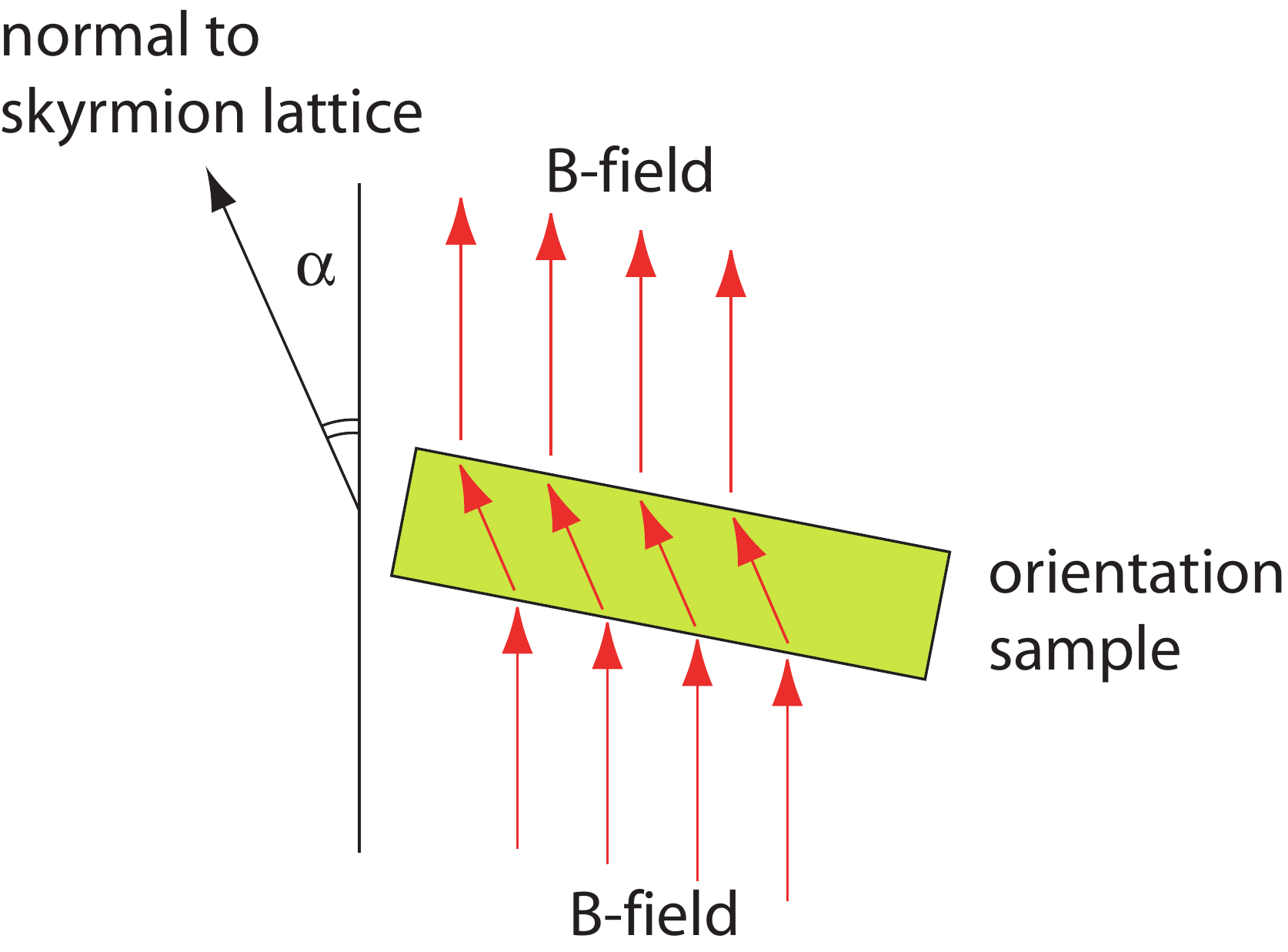}
\end{center}
\caption{Schematic view as seen from above of the variation of the magnetic field lines across the sample. Due to demagnetizing effects at the fringes of the sample the internal field neither coincides with the tilting angle nor with the incident neutron beam.} \label{figure-S2}
\end{figure}

An easy way to notice the influence of the demagnetizing fields was in the context of the orientation of the hexagonal scattering intensity with respect to the orientation of the sample as described in the supplementary online information of Ref.\,\cite{mueh09}. When the applied field was not perfectly perpendicular to the disc, we observed a deflection of the maxima consistent with the demagnetizing fields as illustrated in Fig.~\ref{figure-S2}.

In some of our studies rocking scans could only be carried out for the vertical axis, i.e., along the long side of our sample which was parallel to a $\langle110\rangle$ direction. It was oriented vertical and perpendicular to the incident neutron beam and magnetic field as shown in Fig.~\ref{figureS1}\,(A). Thus in the A-phase two spots of the hexagonal scattering intensity coincided always with the vertical $\langle110\rangle$ axis and thus the axis of the rocking scans.

\begin{figure}
\begin{center}
\includegraphics[width=0.5\textwidth]{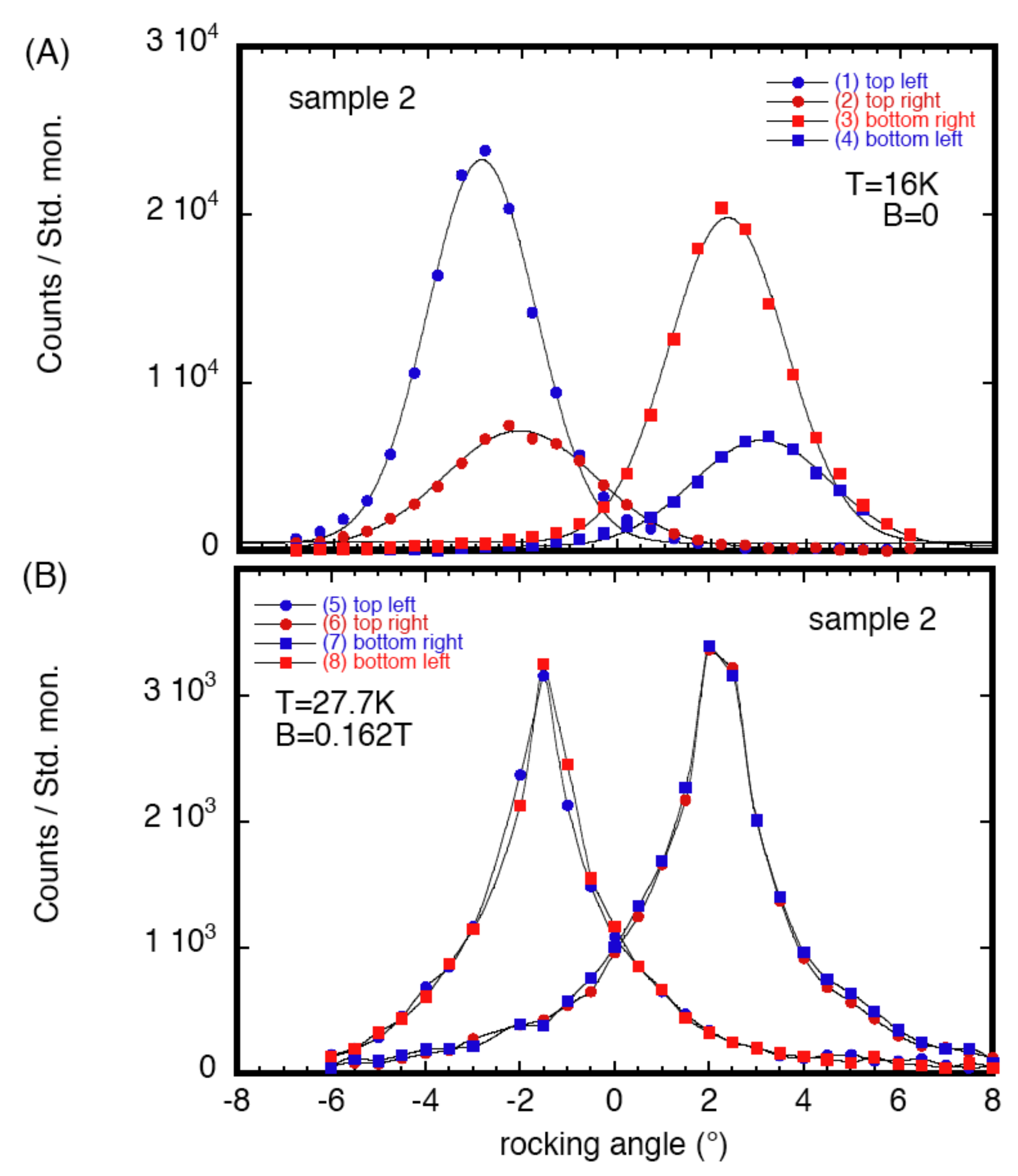}
\end{center}
\caption{(A) Rocking scans in the helical state of MnSi. A Gaussian distribution with a width at half-maximum corresponding to a magnetic mosaicity $\eta_{m}\approx3.5^{\circ}$ is observed consistent with previous studies and the literature \cite{pfle07b}. (B) Rocking scans in the A-phase of MnSi. The apparent width is somewhat reduced, however, the functional dependence is completely different. As shown in Fig.\,\ref{figure-S4} this distribution is most likely due to demagnetizing effects. See also the online supplementary information of Ref.\,\cite{mueh09}.} \label{figure-S3}
\end{figure}

For a reasonably well collimated neutron beam the observation of equal intensity of the vertical spots implies, that neither of these vertical spots satisfies the Bragg condition. The intensity originates only from the tails of the Bragg spot. Therefore, when summing over rocking scans with respect to the vertical axis vertical spots remain weak.

Shown in Fig.~\ref{figure-S3} are the rocking scans for sample (ii) studied in the spin torque measurements. In the helical state the half-width of the rocking scans corresponded to a magnetic mosaicity $\eta_{m}\approx3.5^{\circ}$ consistent with previous work and long range order \cite{pfle07b}. In the A-phase the half-width of the rocking scans corresponded to an apparent magnetic mosaicity $\eta_{m}\approx1.75^{\circ}$.

\begin{figure}
\begin{center}
\includegraphics[width=0.5\textwidth]{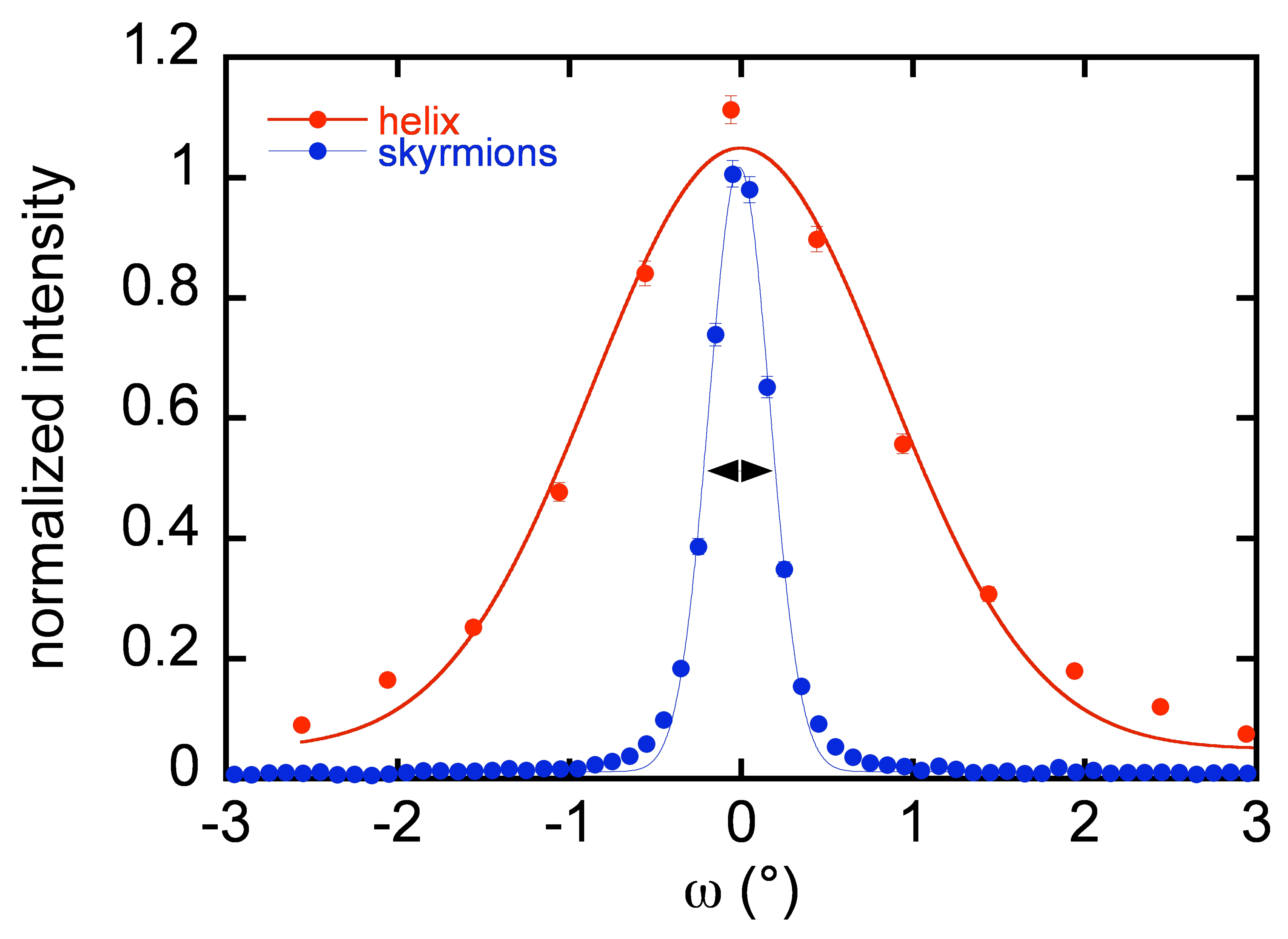}
\end{center}
\caption{Typical intensity variation in rocking scans in the helical state and the skyrmion lattice of MnSi for a sample (v). Only the central section of the sample was illuminated by the neutrons, thereby avoiding any fringe effects at the rim of the sample. This way the internal field distribution was extremely narrow. The rocking scans show a width $\Delta\omega\approx3.5^{\circ}$ at half maximum in excellent agreement with all previous studies and the literature. In contrast, in the skyrmion lattice the rocking width is resolution limited $\Delta\omega\approx0.4^{\circ}$. This suggests strongly that the skyrmion lattice is very well ordered and rather rigid as assumed in the discussion in the main text.}
\label{figure-S4}
\end{figure}

In order to identify the origin of the peculiar angular dependence observed in the skyrmion lattice in the spin torque sample we measured rocking scans of the helical order and the skyrmion lattice in a disc that was only a few tenths of a mm thick with a diameter of roughly $15\,{\rm mm}$. The disc was positioned perpendicular to the magnetic field and incident neutron beam. Using a small source aperture only a tiny cross-section of a few ${\rm mm^2}$ in the center of the disc was studied, thereby avoiding any anomalous internal field distributions at the fringes of the sample.

Typical rocking scans observed in this configuration are shown in Fig.\,\ref{figure-S4}. For convenience the intensity is normalized. In the helical state the rocking width is $\Delta\omega\approx3.5^{\circ}$, in excellent agreement with values reported in the literature \cite{pfle07b}. This value also agrees very well with the rocking width observed in the spin torque sample. Thus the rocking width appears to reflect the intrinsic magnetic mosaicity of the helimagnetic state in MnSi.

In contrast, the rocking width in the skyrmion lattice is now resolution limited with $\Delta\omega\approx0.4^{\circ}$ -- almost an order of magnitude smaller.  Astonishingly, the corresponding coherence length of the magnetic state is of the order of several ten micro-meters! In turn this observation strongly suggests, that the rocking width observed in the spin torque sample shown in Fig.\,\ref{figure-S3}(B) results from the distribution of demagnetizing fields in the sample. In other words, the skyrmion lines closely follow the internal field distribution and the skyrmion lattice exhibits well developed long-range order. This is an important assumption for the analysis of our data in the main text.

\newpage
\section{Theoretical analysis}

The magnetic properties of MnSi are well described by a Ginzburg-Landau theory for the local magnetization $\vec M(\vec{r})$,
\begin{eqnarray} \label{size}
F[\vec M]=\int\!d^3r\,\left( r_0\vec M^2+J(\nabla\vec M)^2+2D\,\vec
M\cdot(\nabla\times\vec M)+U\vec M^4 -\vec B\cdot \vec M\right), \label{F}
\end{eqnarray}
where $\vec B$ is the external
magnetic field and $r_0,D,J,U$ are parameters ($U,J>0$) \cite{naka80,bak80}.
As shown in Ref.\,\cite{mueh09}, this theory describes the formation and phase diagram of the skyrmion lattice if
thermal fluctuations around the mean field solution, i.e. around local minima of (\ref{size}), are taken into account.
The energy scales and the hierarchy of forces in the system are
governed by the strength of spin-orbit coupling, $\lambda_{SO}$,
which is very small in MnSi. The most
important component attributed to spin-orbit coupling is thereby
the Dzyaloshinsky Moriya interaction, $D\propto \lambda_{SO}$,
which determines the size and distance of the skyrmions
given by $J/D \propto 1/\lambda_{SO}$
(for details see Ref.\,\cite{mueh09}).

\subsection{Orientation of the Skyrmion lattice}

The Ginzburg-Landau potential in lowest order in $\lambda_{SO}$ given in  Eq.~(\ref{size}) is rotationally
invariant around the axis defined by the magnetic field $\vec B$. The orientation of the skyrmion lattice within
the plane perpendicular to $\vec B$ in the absence of a current, $j=0$, is determined by the anisotropy of the
atomic crystal, i.e., anisotropy terms of higher order in $\lambda_{SO}$ that are not yet included in
Eq.~(\ref{size}). Such anisotropies favor certain orientation angles $\Phi$ of the magnetic texture. Due to the
sixfold symmetry of the skyrmion lattice, a $\Phi$-potential proportional to at least $-\cos( 6 \Phi)$ is needed
in order to orient it. Such anisotropy terms are, however, suppressed by the tiny factor $\lambda_{SO}^6$ with
$\lambda_{SO} \sim 10^{-2}$. An example of such a term is $\int (\partial_x^3 \vec M)^2+(\partial_y^3 \vec
M)^2+(\partial_z^3 \vec M)^2$ which is proportional to $\lambda_{SO}^6$ 
because the distance between the skyrmions, $\sim J/D$, is linear in $1/\lambda_{SO}$. 
In contrast, the energy density, $e_s$, of the skyrmion lattice, which according to
specific heat measurements \cite{bauer10} is of the
order of $\Delta F/V \sim 10^{-2} k_B T_c/a^3$, 
is only proportional to $\lambda_{SO}^2$ \cite{mueh09}. 
The torque per volume resulting
from such a potential can therefore be estimated as 
$\tau_L \sim \lambda_{SO}^4 \partial_\Phi \cos( 6 \Phi) \Delta F/V$, which
reproduces Eq.~(3) in the main text.

\subsection{Disorder pinning of the Skyrmion lattice}

In our spin torque experiment an unexpectedly small threshold current $j_c \sim 10^{-6}\,{\rm A\,m^{-2}}$ was observed
(Fig.~3(A) of the main text) suggesting an inefficient pinning of the Skyrmion lattice by disorder. It is instructive to compare
this observation with the seemingly similar situation encountered in type II superconductors where the pinning of vortex lines
has been studied extensively \cite{reviewVort94}. In superconductors the order parameter varies on a length scale
typically much smaller than the distance between the vortices and, in particular, vanishes in the vortex cores. A pinning
force $F$ due to an impurity can be estimated by taking the energy difference $\Delta E$ of a defect in the core center
and a defect outside of the superconducting vortex core divided by the typical radius, $r_c$, of the core, $F \sim \Delta E/r_c$.

In contrast, pinning may be expected to differ significantly in the case of a Skyrmion lattice because
the magnetization is smooth and does not vanish anywhere \cite{mueh09}.
Besides the low defect concentration of our samples, evident from the
large charge carrier mean free paths around 1000\,{\AA}, there are several aspects that have to be taken into
account when estimating the pinning force arising from a single, localized defect: (i)
The free energy density  of the skyrmion lattice as estimated from the measured specific
heat \cite{bauer10}, $\Delta F/V \sim 10^{-2} k_B T_c/a^3$, is tiny. (ii) As already mentioned above, pinning
forces arising from the coupling to the magnitude of the magnetization are small because it varies
by less than $\pm20$\% \cite{mueh09} within the magnetic unit cell. (iii) The coupling of disorder to
the direction of the spins originates from spin-orbit interactions. It is hence also very small. (iv) The
forces, i.e., the energy changes {\em per length}, are small because the distance between the skyrmions is a factor 40
larger than $a$. Combining all these factors, we estimate the pinning force from a single strong defect (e.g. by replacing
a Mn atom by a nonmagnetic impurity) to be less than a few $10^{-5} k_B T_c/a$. Here we assumed that
the magnetization is completely destroyed in a fraction of the unit cell
(there are 4 Mn and 4 Si atoms per unit cell) and that the difference of the free energy for a defect in the center
of a skyrmion and, e.g., between skyrmions is only a fraction of $\Delta F/V$ due to the small
variations of the amplitude of the magnetization. The most important factor is, however, the large distance between the skyrmions.

As discussed in the main text, the total pinning force per volume can {\em not} be obtained by multiplying
the pinning force of a single defect by the defect density, because the pinning forces from several defects within a domain
can partially average out if the skyrmion lattice is sufficiently rigid. The analogous collective pinning regime has been
studied extensively for superconducting vortices~\cite{reviewVort94}. For a full quantitative theory of pinning, one
would therefore need information on (i) the nature of single pinning centers, (ii) the defect concentration and (iii) the rigidity
of the skyrmion lattice. While the latter can be calculated or measured with neutron scattering and while
some information of defect concentration can in principle be inferred from the residual resistivity or de-Haas-van-Alphen measurements, the nature of defects important for either magnetic pinning or scattering of electrons is presently not clear.
Nevertheless, we conclude that besides the small defect concentration
the smoothness of the skyrmion spin structure is probably the most important factor
to explain the small pinning forces.

\newpage